\documentclass{webofc}
\usepackage[varg]{txfonts}   
\usepackage{tikz}
\usepackage{verbatim}
\begin{document}

\title{Ordering According to Size of Disks in a Narrow Channel}

\author{\firstname{Dan} \lastname{LIU}\inst{1}\fnsep\thanks{\email{dliu@hartford.edu}} \and
        \firstname{Michael} \lastname{KARBACH}\inst{2}\fnsep\thanks{\email{michael@karbach.org}} \and
        \firstname{Gerhard} \lastname{M\"ULLER}\inst{3}\fnsep\thanks{\email{gmuller@uri.edu}}
}

\institute{Department of Physics, University of Hartford, West Hartford CT 06443, USA 
\and
 Fachgruppe Physik, Bergische Universität Wuppertal, 42119 Wuppertal, Germany
\and
    Department of Physics, University of Rhode Island, Kingston RI 02881, USA
}
\abstract{A long and narrow channel confines disks of two sizes. 
The disks are randomly agitated in a widened channel under moderate pressure, then jammed according to a tunable protocol.
We present exact results that characterize jammed macrostates (volume, entropy, jamming patterns).
The analysis divides jammed disk sequences into overlapping tiles out of which statistically interacting quasiparticles are constructed. 
The fractions of small and large disks are controlled by a chemical potential adapted to configurational statistics of granular matter.
The results show regimes for the energy parameters (determined by the jamming protocol) that either enhance or suppress the mixing of disk sizes. 
Size segregation or size alternation driven by steric forces alone are manifestations of a broken symmetry.}

\maketitle

\section{Introduction}\label{sec-1}

In polydisperse granular matter, grain segregation according to size, weight, or shape is a common phenomenon, desirable or undesirable depending on circumstances.
The most common causes of segregation are a gravitational field and centrifugal forces \cite{PHBJ22}.
The processes of segregation under these two influences have been widely and systematically studied.
Grain segregation can also be caused by container walls of particular shape \cite{ABS22}. 
These causes of segregation have in common that they favor, in one way or another, a direction. 

Are there scenarios in which grain segregation is associated with spontaneous symmetry breaking?
The best chances for providing an affirmative answer present themselves to studies of polydisperse grains in narrow channels, where caging effects permit a systematic characterization of jammed microstates.
Advances made by analytical studies in this line of granular matter research have employed a variety of strategies and reported interesting results that complement simulation data for comparable situations \cite{BS06, AB09, BA11, janac1, janac2, janac3}.

This work builds on a technique that adapts the concept of statistically interacting particles to granular matter \cite{janac1, janac2, janac3}.
The investigation of situations that make grain segregation possible requires two major extensions in the methodology.
(i) The jamming protocol must permit the randomization of disks according to size.
(ii) The grandcanonical ensemble must be adapted to granular statistics along with a quantity akin to a chemical potential \cite{janac3, GZB23}.
This short account outlines the method of analysis and presents a few key results. An account which includes the mathematical analysis is in the works.

\section{Jamming Patterns}\label{sec-2}

Disks of two sizes with diameters $\sigma_\mathrm{L}\geq\sigma_\mathrm{S}$ in arbitrary sequence are being jammed in a channel of width $H$.
Every disk has three points of contact, either with an adjacent disk or with a wall. 
All disks have wall contact and no loose disks exist under jamming if
\begin{align}\label{eq:1} 
1\leq \frac{\sigma_\mathrm{L}}{\sigma_\mathrm{S}}
<\frac{H}{\sigma_\mathrm{S}}<
1+\sqrt{3/4}.
\end{align}
All jammed microstates can be assembled from 16 tiles composed of two adjacent disks with one disk overlapping (Table~\ref{tab:t1}).
Adding a tile to an already existing string of tiles must match the pattern regarding size and position of the overlapping disk and maintain mechanical stability under jamming forces.
Each tile has one of six distinct volumes (Table ~\ref{tab:t2}).

\begin{table}[htb]
\caption{Distinct tiles that constitute jammed microstates of arbitrary disk sequences subject to the constraints (\ref{eq:1}). The ``ID'' tile must be followed one of the ``rule'' tiles.  The motifs shown pertain to $\sigma_\mathrm{L}=2$, $\sigma_\mathrm{S}=1.4$, $H=2.5$.}

  \label{tab:t1}
\begin{center}
\begin{tabular}{cccc|cccc} \hline\hline
motif & ~ID~ & rule & vol.~ & ~motif~ & ~ID~ & rule & vol.  \\ \hline \rule[-2mm]{0mm}{8mm}
\begin{tikzpicture} [scale=0.2]
\draw (0.0,0.0) -- (3.94,0.0) -- (3.94,2.5) -- (0.0,2.5) -- (0,0);
\filldraw [fill=gray, draw=black] (1.0,1.5) circle (1.0);
\filldraw [fill=gray, draw=black] (2.94,1.0) circle (1.0);
\end{tikzpicture}
& 1 & $2,6,10,14$ & $V_\mathrm{a}$ &   
\begin{tikzpicture} [scale=0.2]
\draw (0.0,0.0) -- (4.0,0.0) -- (4.0,2.5) -- (0.0,2.5) -- (0,0);
\filldraw [fill=gray, draw=black] (1,1.5) circle (1.0);
\filldraw [fill=gray, draw=black] (3.0,1.5) circle (1.0);
\end{tikzpicture}
& 9 & $1,5,9$ & $V_\mathrm{d}$  \\ \rule[-2mm]{0mm}{6mm}

\begin{tikzpicture} [scale=0.2]
\draw (0.0,0.0) -- (3.94,0.0) -- (3.94,2.5) -- (0.0,2.5) -- (0,0);
\filldraw [fill=gray, draw=black] (1,1) circle (1.0);
\filldraw [fill=gray, draw=black] (2.94,1.5) circle (1.0);
\end{tikzpicture}
& 2 & $1,5,9,13$ & $V_\mathrm{a}$ &  
\begin{tikzpicture} [scale=0.2]
\draw (0.0,0.0) -- (4.0,0.0) -- (4.0,2.5) -- (0.0,2.5) -- (0,0);
\filldraw [fill=gray, draw=black] (1.0,1.0) circle (1.0);
\filldraw [fill=gray, draw=black] (3.0,1.0) circle (1.0);
\end{tikzpicture}
& 10 & $2,6,10$ & $V_\mathrm{d}$  \\ \rule[-2mm]{0mm}{6mm}

\begin{tikzpicture} [scale=0.2]
\draw (0.0,0.0) -- (2.27,0.0) -- (2.27,2.5) -- (0.0,2.5) -- (0,0);
\filldraw [fill=gray, draw=black] (0.7,1.8) circle (0.7);
\filldraw [fill=gray, draw=black] (1.57,0.7) circle (0.7);
\end{tikzpicture}
& 3 & $4,8,12,16$ & $V_\mathrm{b}$ & 
\begin{tikzpicture} [scale=0.2]
\draw (0.0,0.0) -- (2.8,0.0) -- (2.8,2.5) -- (0.0,2.5) -- (0,0);
\filldraw [fill=gray, draw=black] (0.7,1.8) circle (0.7);
\filldraw [fill=gray, draw=black] (2.1,1.8) circle (0.7);
\end{tikzpicture}
& 11 & $3,7,15,11$ & $V_\mathrm{e}$  \\ \rule[-2mm]{0mm}{6mm}

\begin{tikzpicture} [scale=0.2]
\draw (0.0,0.0) -- (2.27,0.0) -- (2.27,2.5) -- (0.0,2.5) -- (0,0);
\filldraw [fill=gray, draw=black] (0.7,0.7) circle (0.7);
\filldraw [fill=gray, draw=black] (1.57,1.8) circle (0.7);
\end{tikzpicture}
& 4 & $3,7,11,15$ & $V_\mathrm{b}$ & 
\begin{tikzpicture} [scale=0.2]
\draw (0.0,0.0) -- (2.8,0.0) -- (2.8,2.5) -- (0.0,2.5) -- (0,0);
\filldraw [fill=gray, draw=black] (0.7,0.7) circle (0.7);
\filldraw [fill=gray, draw=black] (2.1,0.7) circle (0.7);
\end{tikzpicture}
& 12 & $4,8,16,12$ & $V_\mathrm{e}$ \\ \rule[-2mm]{0mm}{6mm}

\begin{tikzpicture} [scale=0.2]
\draw (0.0,0.0) -- (3.2,0.0) -- (3.2,2.5) -- (0.0,2.5) -- (0,0);
\filldraw [fill=gray, draw=black] (1,1.5) circle (1.0);
\filldraw [fill=gray, draw=black] (2.5,0.7) circle (0.7);
\end{tikzpicture}
& 5 & $4,8,12,16$ & $V_\mathrm{c}$ &   
\begin{tikzpicture} [scale=0.2]
\draw (0.0,0.0) -- (3.37,0.0) -- (3.37,2.5) -- (0.0,2.5) -- (0,0);
\filldraw [fill=gray, draw=black] (1.0,1.5) circle (1.0);
\filldraw [fill=gray, draw=black] (2.67,1.8) circle (0.7);
\end{tikzpicture}
& 13 & $3,7,15,11$ & $V_\mathrm{f}$  \\ \rule[-2mm]{0mm}{6mm}

\begin{tikzpicture} [scale=0.2]
\draw (0.0,0.0) -- (3.2,0.0) -- (3.2,2.5) -- (0.0,2.5) -- (0,0);
\filldraw [fill=gray, draw=black] (1,1) circle (1.0);
\filldraw [fill=gray, draw=black] (2.5,1.8) circle (0.7);
\end{tikzpicture}
& 6 & $3,7,11,15$ & $V_\mathrm{c}$ &  
\begin{tikzpicture} [scale=0.2]
\draw (0.0,0.0) -- (3.37,0.0) -- (3.37,2.5) -- (0.0,2.5) -- (0,0);
\filldraw [fill=gray, draw=black] (1.0,1.0) circle (1.0);
\filldraw [fill=gray, draw=black] (2.677,0.7) circle (0.7);
\end{tikzpicture}
& 14 & $4,8,16,12$ & $V_\mathrm{f}$  \\ \rule[-2mm]{0mm}{6mm}

\begin{tikzpicture} [scale=0.2]
\draw (0.0,0.0) -- (3.2,0.0) -- (3.2,2.5) -- (0.0,2.5) -- (0,0);
\filldraw [fill=gray, draw=black] (0.7,1.8) circle (0.7);
\filldraw [fill=gray, draw=black] (2.2,1.0) circle (1.0);
\end{tikzpicture}
& 7 & $2,6,10,14$ & $V_\mathrm{c}$ & 
\begin{tikzpicture} [scale=0.2]
\draw (0.0,0.0) -- (3.37,0.0) -- (3.37,2.5) -- (0.0,2.5) -- (0,0);
\filldraw [fill=gray, draw=black] (0.7,1.8) circle (0.7);
\filldraw [fill=gray, draw=black] (2.37,1.5) circle (1.0);
\end{tikzpicture}
& 15 & $1,5$& $V_\mathrm{f}$  \\ \rule[-2mm]{0mm}{6mm}

\begin{tikzpicture} [scale=0.2]
\draw (0.0,0.0) -- (3.2,0.0) -- (3.2,2.5) -- (0.0,2.5) -- (0,0);
\filldraw [fill=gray, draw=black] (0.7,0.7) circle (0.7);
\filldraw [fill=gray, draw=black] (2.2,1.5) circle (1.0);
\end{tikzpicture}
& 8 & $1,5,9,13$ & $V_\mathrm{c}$ & 
\begin{tikzpicture} [scale=0.2]
\draw (0.0,0.0) -- (3.37,0.0) -- (3.37,2.5) -- (0.0,2.5) -- (0,0);
\filldraw [fill=gray, draw=black] (0.7,0.7) circle (0.7);
\filldraw [fill=gray, draw=black] (2.37,1.0) circle (1.0);
\end{tikzpicture}
& 16 & $2,6$ & $V_\mathrm{f}$ 
\\ \hline\hline
\end{tabular}
\end{center}
\end{table}

\begin{table}[t]
  \caption{Volume of tiles in a channel has unit cross sections. The bracketed portion overlaps in successive tiles and does not contribute the excess volume of quasiparticles.
  The numerical values pertain to $\sigma_\mathrm{L}=2$, $\sigma_\mathrm{S}=1.4$, $H=2.5$.}\label{tab:t2}
\begin{center}
\begin{tabular}{l|l|l} \hline\hline
 & ~$V=[~~]+\tilde{V}$ & ~$\tilde{V}$  \\ \hline \rule[-2mm]{0mm}{6mm}
$V_\mathrm{a}$~ & ~$[\sigma_\mathrm{L}]+\sqrt{H(2\sigma_\mathrm{L}-H)}$ & ~1.963 
\\ \rule[-2mm]{0mm}{6mm}
$V_\mathrm{b}$~ & ~$[\sigma_\mathrm{S}]+\sqrt{H(2\sigma_\mathrm{S}-H)}$ & ~0.886 
\\ \rule[-2mm]{0mm}{6mm}
$V_\mathrm{c}$~ & ~$[\frac{1}{2}(\sigma_\mathrm{L}+\sigma_\mathrm{S})]
+\sqrt{H(\sigma_\mathrm{L}+\sigma_\mathrm{S}-H)}$~~ & ~1.5 
\\ \rule[-2mm]{0mm}{6mm}
$V_\mathrm{d}$~ & ~$[\sigma_\mathrm{L}]+\sigma_\mathrm{L}$ & ~2.0 
\\ \rule[-2mm]{0mm}{6mm}
$V_\mathrm{e}$~ & ~$[\sigma_\mathrm{S}]+\sigma_\mathrm{S}$ & ~1.4 
\\ \rule[-2mm]{0mm}{6mm}
$V_\mathrm{f}$~ & ~$[\frac{1}{2}(\sigma_\mathrm{L}+\sigma_\mathrm{S})]
+\sqrt{\sigma_\mathrm{L}\sigma_\mathrm{S}}$ & ~1.673 
\\ \hline\hline
\end{tabular}
\end{center}
\end{table}

Microstates of jammed disks are described as configurations of statistically interacting particles generated from the reference state (pseudo-vacuum) containing only large disks and constructed from tiles 1 and 2 exclusively:
\begin{equation}\label{eq:2} 
\mathsf{pv}=\mathsf{1212\cdots 1}\qquad 
\begin{tikzpicture} [scale=0.2]
\draw (0,0) -- (10,0.0);
\draw (0,0) -- (0,2.5);
\draw (0,2.5) -- (10,2.5);
\filldraw [fill=gray, draw=black] (1.0,1.5) circle (1.0);
\filldraw [fill=gray, draw=black] (2.94,1.0) circle (1.0);
\filldraw [fill=gray, draw=black] (4.88,1.5) circle (1.0);
\filldraw [fill=gray, draw=black] (6.82,1.0) circle (1.0);
\filldraw [fill=gray, draw=black] (8.74,1.5) circle (1.0);
\filldraw [fill=gray, draw=black] (11,1.25) circle (0.1);
\filldraw [fill=gray, draw=black] (12,1.25) circle (0.1);
\filldraw [fill=gray, draw=black] (13,1.25) circle (0.1);
\end{tikzpicture}\hspace{2mm}
\begin{tikzpicture} [scale=0.2]
\draw (0.0,0.0) -- (4.14,0.0) -- (4.14,2.5) -- (0.0,2.5);
\filldraw [fill=gray, draw=black] (1.2,1.5) circle (1.0);
\filldraw [fill=gray, draw=black] (3.14,1.0) circle (1.0);
\end{tikzpicture}
\end{equation}
All other jammed microstates can be generated by the activation of quasiparticles from this reference state. 
We have identified $M=17$ species of particles that serve this purpose (Table~\ref{tab:t3}).
The \emph{compacts} $m=1,2$ and \emph{hosts} $m=3,\ldots,14$ modify the pseudo-vacuum (\ref{eq:2}), whereas the \emph{tags} $m=15,16,17$ modify any one of the hosts. 

\begin{table}[t]
  \caption{Quasiparticles activated (directly or indirectly) from reference state (\ref{eq:2}). The excess volumes $\Delta V_m$ are relative to a segment of $\mathsf{pv}$ with the same number of disks. Species $m=7,8,13,14,17$ have two distinct motifs.}\label{tab:t3}
\begin{center}
\begin{tabular}{rl|rl|l} \hline\hline
$m$~ & motif & ~$m$~ & motif & ~$\Delta V_m$ \\ \hline \rule[-2mm]{0mm}{8mm}
1~ & \begin{tikzpicture} [scale=0.2]
\draw (0.0,0.0) -- (4.0,0.0) -- (4.0,2.5) -- (0.0,2.5) -- (0,0);
\filldraw [fill=gray, draw=black] (1,1.5) circle (1.0);
\filldraw [fill=gray, draw=black] (3.0,1.5) circle (1.0);
\end{tikzpicture}
& 
2~ & \begin{tikzpicture} [scale=0.2]
\draw (0.0,0.0) -- (4.0,0.0) -- (4.0,2.5) -- (0.0,2.5) -- (0,0);
\filldraw [fill=gray, draw=black] (1.0,1.0) circle (1.0);
\filldraw [fill=gray, draw=black] (3.0,1.0) circle (1.0);
\end{tikzpicture}
& ~$\tilde{V}_\mathrm{d}-\tilde{V}_\mathrm{a}$
\\ \rule[-2mm]{0mm}{6mm}
3~ & \begin{tikzpicture} [scale=0.2]
\draw (0.0,0.0) -- (5.0,0.0) -- (5.0,2.5) -- (0.0,2.5) -- (0,0);
\filldraw [fill=gray, draw=black] (1,1.5) circle (1.0);
\filldraw [fill=gray, draw=black] (2.5,0.7) circle (0.7);
\filldraw [fill=gray, draw=black] (4.0,1.5) circle (1.0);
\end{tikzpicture}
& 
4~ & \begin{tikzpicture} [scale=0.2]
\draw (0.0,0.0) -- (5.0,0.0) -- (5.0,2.5) -- (0.0,2.5) -- (0,0);
\filldraw [fill=gray, draw=black] (1,1.0) circle (1.0);
\filldraw [fill=gray, draw=black] (2.5,1.8) circle (0.7);
\filldraw [fill=gray, draw=black] (4.0,1.0) circle (1.0);
\end{tikzpicture}
& ~$2\tilde{V}_\mathrm{c}-2\tilde{V}_\mathrm{a}$
\\ \rule[-2mm]{0mm}{6mm}
5~ & \begin{tikzpicture} [scale=0.2]
\draw (0.0,0.0) -- (5.4,0.0) -- (5.4,2.5) -- (0.0,2.5) -- (0,0);
\filldraw [fill=gray, draw=black] (1.0,1.5) circle (1.0);
\filldraw [fill=gray, draw=black] (2.7,1.8) circle (0.7);
\filldraw [fill=gray, draw=black] (4.4,1.5) circle (1.0);
\end{tikzpicture}
& 
6~ & \begin{tikzpicture} [scale=0.2]
\draw (0.0,0.0) -- (5.4,0.0) -- (5.4,2.5) -- (0.0,2.5) -- (0,0);
\filldraw [fill=gray, draw=black] (1.0,1.0) circle (1.0);
\filldraw [fill=gray, draw=black] (2.7,0.7) circle (0.7);
\filldraw [fill=gray, draw=black] (4.4,1.0) circle (1.0);
\end{tikzpicture}
& ~$2\tilde{V}_\mathrm{f}-2\tilde{V}_\mathrm{a}$
\\ \rule[-2mm]{0mm}{6mm}
7~ & \begin{tikzpicture} [scale=0.2]
\draw (0.0,0.0) -- (5.2,0.0) -- (5.2,2.5) -- (0.0,2.5) -- (0,0);
\filldraw [fill=gray, draw=black] (1.0,1.0) circle (1.0);
\filldraw [fill=gray, draw=black] (2.5,1.8) circle (0.7);
\filldraw [fill=gray, draw=black] (4.2,1.5) circle (1.0);
\end{tikzpicture}
&  
7~ & \begin{tikzpicture} [scale=0.2]
\draw (0.0,0.0) -- (5.2,0.0) -- (5.2,2.5) -- (0.0,2.5) -- (0,0);
\filldraw [fill=gray, draw=black] (1.0,1.5) circle (1.0);
\filldraw [fill=gray, draw=black] (2.7,1.8) circle (0.7);
\filldraw [fill=gray, draw=black] (4.2,1.0) circle (1.0);
\end{tikzpicture}
& ~$\tilde{V}_\mathrm{c}+\tilde{V}_\mathrm{f}-2\tilde{V}_\mathrm{a}$
\\ \rule[-2mm]{0mm}{6mm}
8~ & \begin{tikzpicture} [scale=0.2]
\draw (0.0,0.0) -- (5.2,0.0) -- (5.2,2.5) -- (0.0,2.5) -- (0,0);
\filldraw [fill=gray, draw=black] (1.0,1.5) circle (1.0);
\filldraw [fill=gray, draw=black] (2.5,0.7) circle (0.7);
\filldraw [fill=gray, draw=black] (4.2,1.0) circle (1.0);
\end{tikzpicture}
&  
8~ & \begin{tikzpicture} [scale=0.2]
\draw (0.0,0.0) -- (5.2,0.0) -- (5.2,2.5) -- (0.0,2.5) -- (0,0);
\filldraw [fill=gray, draw=black] (1.0,1.0) circle (1.0);
\filldraw [fill=gray, draw=black] (2.7,0.7) circle (0.7);
\filldraw [fill=gray, draw=black] (4.2,1.5) circle (1.0);
\end{tikzpicture}
& ~$\tilde{V}_\mathrm{c}+\tilde{V}_\mathrm{f}-2\tilde{V}_\mathrm{a}$
\\ \rule[-2mm]{0mm}{6mm}
9~ & \begin{tikzpicture} [scale=0.2]
\draw (0.0,0.0) -- (6.07,0.0) -- (6.07,2.5) -- (0.0,2.5) -- (0,0);
\filldraw [fill=gray, draw=black] (1.0,1.5) circle (1.0);
\filldraw [fill=gray, draw=black] (2.7,1.8) circle (0.7);
\filldraw [fill=gray, draw=black] (3.57,0.7) circle (0.7);
\filldraw [fill=gray, draw=black] (5.07,1.5) circle (1.0);
\end{tikzpicture}
&  
10~ & \begin{tikzpicture} [scale=0.2]
\draw (0.0,0.0) -- (6.07,0.0) -- (6.07,2.5) -- (0.0,2.5) -- (0,0);
\filldraw [fill=gray, draw=black] (1.0,1.0) circle (1.0);
\filldraw [fill=gray, draw=black] (2.5,1.8) circle (0.7);
\filldraw [fill=gray, draw=black] (3.37,0.7) circle (0.7);
\filldraw [fill=gray, draw=black] (5.07,1.0) circle (1.0);
\end{tikzpicture}
&  ~$\tilde{V}_\mathrm{b}+\tilde{V}_\mathrm{c}+\tilde{V}_\mathrm{f}-3\tilde{V}_\mathrm{a}$
\\ \rule[-2mm]{0mm}{6mm}
11~ & \begin{tikzpicture} [scale=0.2]
\draw (0.0,0.0) -- (6.07,0.0) -- (6.07,2.5) -- (0.0,2.5) -- (0,0);
\filldraw [fill=gray, draw=black] (1.0,1.5) circle (1.0);
\filldraw [fill=gray, draw=black] (2.5,0.7) circle (0.7);
\filldraw [fill=gray, draw=black] (3.37,1.8) circle (0.7);
\filldraw [fill=gray, draw=black] (5.07,1.5) circle (1.0);
\end{tikzpicture}
&  
12~ & \begin{tikzpicture} [scale=0.2]
\draw (0.0,0.0) -- (6.07,0.0) -- (6.07,2.5) -- (0.0,2.5) -- (0,0);
\filldraw [fill=gray, draw=black] (1.0,1.0) circle (1.0);
\filldraw [fill=gray, draw=black] (2.7,0.7) circle (0.7);
\filldraw [fill=gray, draw=black] (3.57,1.8) circle (0.7);
\filldraw [fill=gray, draw=black] (5.07,1.0) circle (1.0);
\end{tikzpicture}
&  ~$\tilde{V}_\mathrm{b}+\tilde{V}_\mathrm{c}+\tilde{V}_\mathrm{f}-3\tilde{V}_\mathrm{a}$
\\ \rule[-2mm]{0mm}{6mm}
13~ & \begin{tikzpicture} [scale=0.2]
\draw (0.0,0.0) -- (6.27,0.0) -- (6.27,2.5) -- (0.0,2.5) -- (0,0);
\filldraw [fill=gray, draw=black] (1.0,1.5) circle (1.0);
\filldraw [fill=gray, draw=black] (2.7,1.8) circle (0.7);
\filldraw [fill=gray, draw=black] (3.57,0.7) circle (0.7);
\filldraw [fill=gray, draw=black] (5.27,1.0) circle (1.0);
\end{tikzpicture}
&  
13~ & \begin{tikzpicture} [scale=0.2]
\draw (0.0,0.0) -- (6.27,0.0) -- (6.27,2.5) -- (0.0,2.5) -- (0,0);
\filldraw [fill=gray, draw=black] (1.0,1.0) circle (1.0);
\filldraw [fill=gray, draw=black] (2.7,0.7) circle (0.7);
\filldraw [fill=gray, draw=black] (3.57,1.8) circle (0.7);
\filldraw [fill=gray, draw=black] (5.27,1.5) circle (1.0);
\end{tikzpicture}
& ~$\tilde{V}_\mathrm{b}+2\tilde{V}_\mathrm{f}-3\tilde{V}_\mathrm{a}$
\\ \rule[-2mm]{0mm}{6mm}
14~ & \begin{tikzpicture} [scale=0.2]
\draw (0.0,0.0) -- (5.87,0.0) -- (5.87,2.5) -- (0.0,2.5) -- (0,0);
\filldraw [fill=gray, draw=black] (1.0,1.0) circle (1.0);
\filldraw [fill=gray, draw=black] (2.5,1.8) circle (0.7);
\filldraw [fill=gray, draw=black] (3.37,0.7) circle (0.7);
\filldraw [fill=gray, draw=black] (4.87,1.5) circle (1.0);
\end{tikzpicture}
& 
14~ & \begin{tikzpicture} [scale=0.2]
\draw (0.0,0.0) -- (5.87,0.0) -- (5.87,2.5) -- (0.0,2.5) -- (0,0);
\filldraw [fill=gray, draw=black] (1.0,1.5) circle (1.0);
\filldraw [fill=gray, draw=black] (2.5,0.7) circle (0.7);
\filldraw [fill=gray, draw=black] (3.37,1.8) circle (0.7);
\filldraw [fill=gray, draw=black] (4.87,1.0) circle (1.0);
\end{tikzpicture}
& ~$\tilde{V}_\mathrm{b}+2\tilde{V}_\mathrm{c}-3\tilde{V}_\mathrm{a}$
\\ \rule[-2mm]{0mm}{6mm}
15~ & \begin{tikzpicture} [scale=0.2]
\draw (0.0,0.0) -- (2.8,0.0) -- (2.8,2.5) -- (0.0,2.5) -- (0,0);
\filldraw [fill=gray, draw=black] (0.7,1.8) circle (0.7);
\filldraw [fill=gray, draw=black] (2.1,1.8) circle (0.7);
\end{tikzpicture}
&  
16~ & \begin{tikzpicture} [scale=0.2]
\draw (0.0,0.0) -- (2.8,0.0) -- (2.8,2.5) -- (0.0,2.5) -- (0,0);
\filldraw [fill=gray, draw=black] (0.7,0.7) circle (0.7);
\filldraw [fill=gray, draw=black] (2.1,0.7) circle (0.7);
\end{tikzpicture}
& ~$\tilde{V}_\mathrm{e}-\tilde{V}_\mathrm{a}$
\\ \rule[-2mm]{0mm}{6mm}
17~ & \begin{tikzpicture} [scale=0.2]
\draw (0.0,0.0) -- (3.14,0.0) -- (3.14,2.5) -- (0.0,2.5) -- (0,0);
\filldraw [fill=gray, draw=black] (0.7,0.7) circle (0.7);
\filldraw [fill=gray, draw=black] (1.57,1.8) circle (0.7);
\filldraw [fill=gray, draw=black] (2.44,0.7) circle (0.7);
\end{tikzpicture}
& 
17~ & \begin{tikzpicture} [scale=0.2]
\draw (0.0,0.0) -- (3.14,0.0) -- (3.14,2.5) -- (0.0,2.5) -- (0,0);
\filldraw [fill=gray, draw=black] (0.7,1.8) circle (0.7);
\filldraw [fill=gray, draw=black] (1.57,0.7) circle (0.7);
\filldraw [fill=gray, draw=black] (2.44,1.8) circle (0.7);
\end{tikzpicture}
&  ~$2\tilde{V}_\mathrm{b}-2\tilde{V}_\mathrm{a}$
\\ \hline\hline
\end{tabular}
\end{center}
\end{table}

\section{Combinatorics and Energetics}\label{sec-3}

The particles identified in Table~\ref{tab:t3} are statistically interacting in the sense that activating one particle from any species $n$ affects the number $d_m$ of remaining slots for the activation of further particles from each species $m$.
This type of interaction can be accounted for by a generalized Pauli principle, here rendered in integrated form \cite{Hald91a, Wu94, janac1},
\begin{equation}\label{eq:5}
  d_m =A_m-\sum_{n=1}^M g_{mn}(N_{n}-\delta_{mn}).
\end{equation}
The capacity constants $A_m$ and interaction coefficients $g_{mn}$ were determined in a different application of the same combinatorics \cite{pichs}.
In a population of $N_n$ particles from all species $n$ already present, Eq.~(\ref{eq:5}) states that there are $d_m$ ways of placing a particle of species $m$.

The reference state (\ref{eq:2}) contains only large disks and the activation of particles from species $m=1,2$ keeps it that way. 
Particles from all other species replace one or two large disks by small disks.
This substitution is encoded in the quantum number $s_m$ for later use:
\begin{equation}\label{eq:3} 
s_m=\left\{\begin{array}{ll}
0 & :~ m=1,2, \\
1 & :~ m=3,4,5,6,7,8,1,15, \\
2 & :~ m=9,10,11,12,13,14,17. \end{array} \right.
\end{equation} 
The number of jammed microstates with particle content $\{N_m\}$ becomes \cite{Hald91a, Wu94, janac1}:
\begin{equation}\label{eq:6} 
W(\{N_m\})=\prod_{m=1}^M\left(\begin{array}{c}d_m+N_m-1 \\ N_m\end{array}\right). 
\end{equation}

The statistical weight of jammed microstates is governed by two kinds of energy present prior to jamming, which, in turn, are controlled by the jamming protocol.
\begin{itemize}

\item[--] Kinetic energy associated with random agitations of controllable intensity $T_\mathrm{k}$, the granular-system equivalent of the thermal energy $k_\mathrm{B}T$, such as applicable to colloids.

\item[--] Potential energy associated with work against the pistons at the ends of the channel. 
This form of energy is encoded in a set of distinct activation energies $\epsilon_m$ for each particle species.

\end{itemize}
The $\epsilon_m$ at given $T_\mathrm{k}$ govern the propensity of specific disk configurations to be realized in jammed microstates.
Each microstate is assigned an energy expression in the form, 
\begin{equation}\label{eq:4} 
E(\{N_m\})=E_\mathrm{pv}+\sum_{m=1}^MN_m\epsilon_m.
\end{equation}

\section{Configurational Statistics}\label{sec-4}

The analysis of statistically interacting particles was adapted in Refs.~\cite{janac1, janac2, janac3} to the configurational statistical analysis of jammed disks.
The granular equivalent of thermodynamic limit implemented at the most fundamental level  concerns the capacity constants and affects primarily the average particle content of jammed macrostates:
\begin{equation}\label{eq:13}
\bar{A}_m\doteq\lim_{N\to\infty}\frac{A_m}{N}, \quad 
\bar{N}_m\doteq\lim_{N\to\infty}\frac{\langle N_m\rangle}{N}.
\end{equation}
From a physics perspective, two quantities of interest, expressed as functions of particle content, are the excess volume and the configurational entropy:
\begin{equation}\label{eq:7} 
\bar{V}\doteq\lim_{N\to\infty}\frac{V-V_\mathrm{pv}}{N}=\sum_{m=1}^M\bar{N}_m\Delta V_m,
\end{equation}
\begin{align}\label{eq:20}
\bar{S} = \lim_{N\to\infty}\frac{S}{k_B}
&=\sum_{m=1}^M\Big[\big(\bar{N}_{m}
+\bar{Y}_m\big)\ln\big(\bar{N}_m+\bar{Y}_m\big) 
  -\bar{N}_m \ln \bar{N}_m -\bar{Y}_m\ln \bar{Y}_m\Big],
 \nonumber \\ 
\bar{Y}_m &\doteq \bar{A}_m-\sum_{n=1}^Mg_{mn} \bar{N}_{n}.
\end{align}
Important for the mathematical analysis is the invertibility of the (asymmetric) matrix,
\begin{equation}\label{eq:14}
G_{mn}=g_{mn}+w_m\delta_{mn}.
\end{equation}
It delivers the population densities of particles in jammed macrostates as the solution of a set of linear equations,
\begin{equation}\label{eq:15}
\sum_nG_{mn}\bar{N}_n=\bar{A}_m\quad \Rightarrow~
\bar{N}_n=\sum_mG^{-1}_{nm}\bar{A}_m,
\end{equation}
on the condition that we know the variables $w_m$, which are the physically relevant solutions of the set of nonlinear algebraic equations \cite{Wu94,janac1},
\begin{equation}\label{eq:16} 
e^{\beta\epsilon_m}=(1+w_m)\prod_{n=1}^M \big(1+w_{n}^{-1}\big)^{-g_{nm}}, \quad
\beta\doteq\frac{1}{T_\mathrm{k}}.
\end{equation}
The granular equivalent of the grand partition function and the grand potential,
\begin{equation}\label{eq:18}
\bar{Z}=\prod_{m=1}^M\left(\frac{1+w_m}{w_m}\right)^{\bar{A}_m},\quad \bar{\Omega}=-\beta^{-1}\ln\bar{Z},
\end{equation}
encapsulate the information needed for the physical quantities of interest.

Experimentally, the channel contains fixed numbers of large and small disks. 
This corresponds to a canonical ensemble. 
However, our methodology operates in the grandcanonical ensemble, where average numbers of large and small disks are determined by the activation energies $\epsilon_m$.
We can control the fractions $\bar{N}_\mathrm{S}$ and $\bar{N}_\mathrm{L}=1-\bar{N}_\mathrm{S}$ of small and large disks, respectively, by an amended activation energy,
\begin{equation}\label{eq:12}
\epsilon_m=p_m-\mu s_m,
\end{equation}
where $p_m$ represents work against the pressure of the pistons exerted prior to jamming.
The quantum number $s_m$, as stated in (\ref{eq:3}), counts the number of small disks in particles of species $m$ activated from reference state (\ref{eq:2}).
The partition function (\ref{eq:18}) with control variables $\beta$ and $\mu$ produces a unique functional relation, 
\begin{equation}\label{eq:8}
\bar{N}_\mathrm{S}(\beta,\mu)=\sum_{m=1}^Ms_m\bar{N}_m,
\end{equation}
which allows us to keep $\bar{N}_\mathrm{S}$ fixed for any value of $\beta$, independent of the $p_m$ selected by a specific jamming protocol.

\section{Activation energies}\label{sec-5}

To this point, the analytic solution (merely outlined here) is exact within the framework of configurational statistics.
It can be completed on this level of rigor for arbitrary activation energies (\ref{eq:12}).
However, physically meaningful results require specific values for the pressure work contributions $p_m$ and for the intensity $T_\mathrm{k}=\beta^{-1}$ of random agitations.
It is this part of the modeling where additional assumptions come into play.

The jamming protocol combines two components.
Component (A) increases the pressure from a moderate value $P$ to a much higher value that stops all motion.
Component (B) narrows the channel from a width that allows the exchange of disk positions in the sequence down to the jamming width $H$.
In combination, (A) and (B) freeze out jammed configuration of large and small disks with probabilities depending on the specific implementation.

The default choice for the first term in Eq.~(\ref{eq:12}) is
\begin{equation}\label{eq:39}
p_m=P\Delta V_m,
\end{equation}
where the $\Delta V_m$ from Table~\ref{tab:t3} are attributes of the jammed microstate.
The effects of different jamming protocols on the selection of jammed configurations can be accounted for by variations in the parameters $\Delta V_m$ from their default values.
Irrespective of how the $\Delta V_m$ are being selected and modified, it is convenient to set $P=1$, which converts excess volumes into measures of expansion work against the piston prior to jamming.

\section{Steric Ordering Tendencies}\label{sec-6}

It turns out that there are two regimes with distinct ordering tendencies, which only depend on a 1-parameter variation in the $\Delta V_m$ involving two large disks and two small disks.
This critical variation pertains to the following two configurations:
\begin{equation}\label{eq:30}
\begin{tikzpicture} [scale=0.2]
\draw (0,0) -- (0,2.5);
\draw (0,0) -- (6.2,0);
\draw (0,2.5) -- (6.2,2.5);
\draw (6.2,0) -- (6.2,2.5);
\filldraw [fill=gray, draw=black] (1,1.5) circle (1.0);
\filldraw [fill=gray, draw=black] (2.5,0.7) circle (0.7);
\filldraw [fill=gray, draw=black] (4.0,1.5) circle (1.0);
\filldraw [fill=gray, draw=black] (5.5,0.7) circle (0.7);
\end{tikzpicture} ~\rightarrow~ 2\tilde{V}_\mathrm{c},\qquad 
\begin{tikzpicture} [scale=0.2]
\draw (0,0) -- (0,2.5);
\draw (0,0) -- (6.01,0);
\draw (0,2.5) -- (6.01,2.5);
\draw (6.01,0) -- (6.01,2.5);
\filldraw [fill=gray, draw=black] (1.0,1.5) circle (1.0);
\filldraw [fill=gray, draw=black] (2.94,1.0) circle (1.0);
\filldraw [fill=gray, draw=black] (4.44,1.8) circle (0.7);
\filldraw [fill=gray, draw=black] (5.31,0.7) circle (0.7);
\end{tikzpicture} ~\rightarrow~ \tilde{V}_\mathrm{a}+\tilde{V}_\mathrm{b},
\end{equation}
The parameter on which the ordering tendency pivots is
\begin{equation}\label{eq:31}
 \Delta\mathcal{V}\doteq 2\tilde{V}_\mathrm{c}-(\tilde{V}_\mathrm{a}+\tilde{V}_\mathrm{b}).
\end{equation}

If we employ the default activation energies, we have $\Delta\mathcal{V}>0$ and size segregation becomes the ordering tendency in the exact analysis as outlined.
However, different jamming protocols are likely associated with activation energies for which $\Delta\mathcal{V}<0$ is the best match, in which case the ordering tendency switches to size alternation in the same exact analysis.

Jamming favors compact microstates.
The two distinct ordering tendencies (segregation or alternation) are dominated by tiles 1 to 8 from Table~\ref{tab:t1}. 
These tiles form domains of compact configurations for given $0<\bar{N}_\mathrm{S}<1$.
Domains of size-segregated disks are sequences of tiles 1 to 4 and domains in a size-alternating pattern are sequences of tiles 5 to 8. 
The role of tiles 9 to 16 is reduced to (less compact) walls between ordered domains. 

If the jamming protocol calls for $\Delta\mathcal{V}>0$, every pair of large disks combined with a pair of small disks is energetically favored in the configuration on the right in (\ref{eq:30}). 
The jammed state frozen out of the lowest-energy unjammed state has all disks size-segregated even if one size is in the majority.
On the other hand, if the jamming protocol calls for $\Delta\mathcal{V}<0$, the energetically favorable configuration of two small and two large disks is the one on the left in (\ref{eq:30}).
The jammed state frozen out of the lowest-energy unjammed state will exhibit an alternating pattern of a length determined by the fraction of disks of minority size.

\section{Segregation, Alternation, or Randomness According to Size}\label{sec-7}

\begin{figure}[h!]
  \begin{center}
\includegraphics[width=50mm]{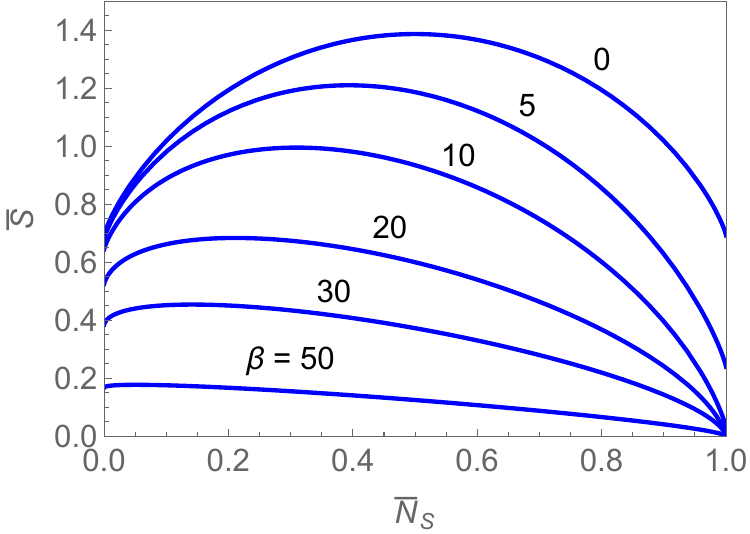}%
\hspace*{5mm}\includegraphics[width=50mm]{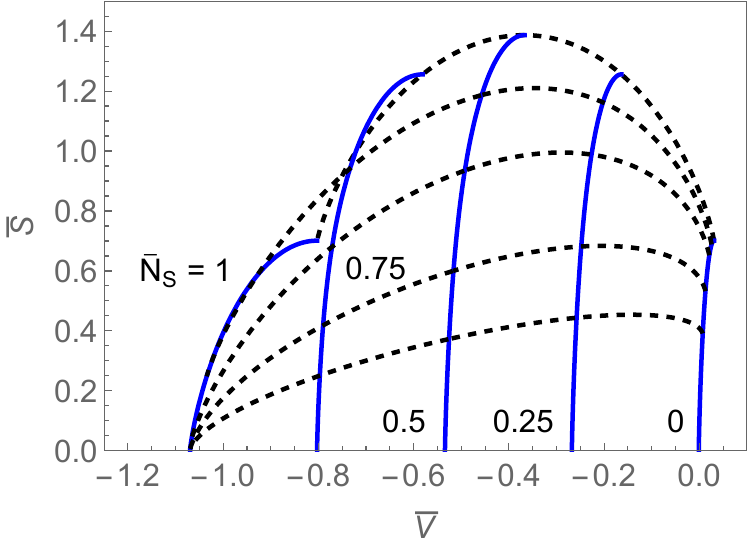}
\includegraphics[width=50mm]{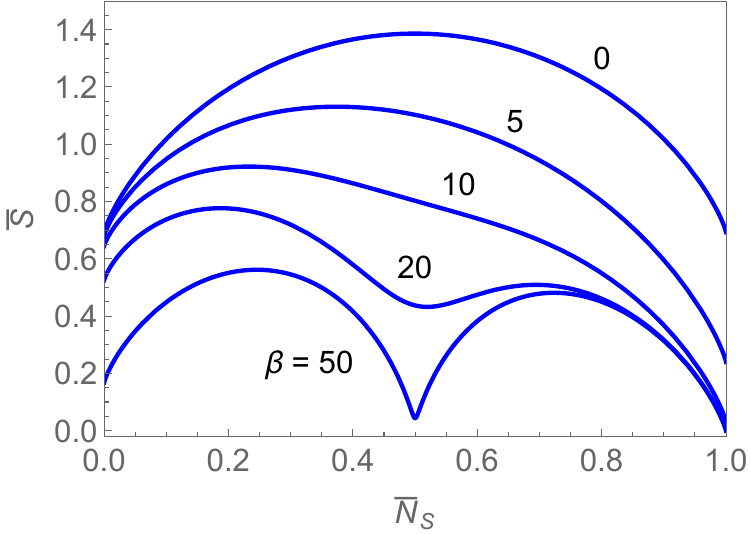}%
\hspace*{5mm}\includegraphics[width=50mm]{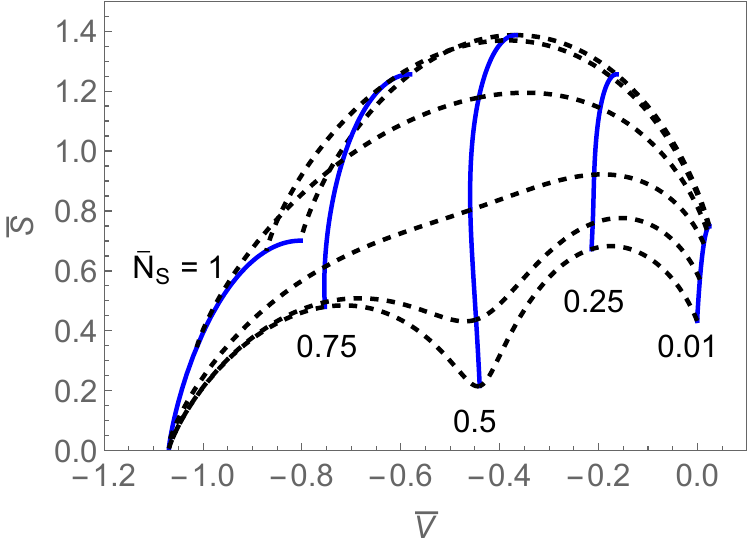}
\includegraphics[width=50mm]{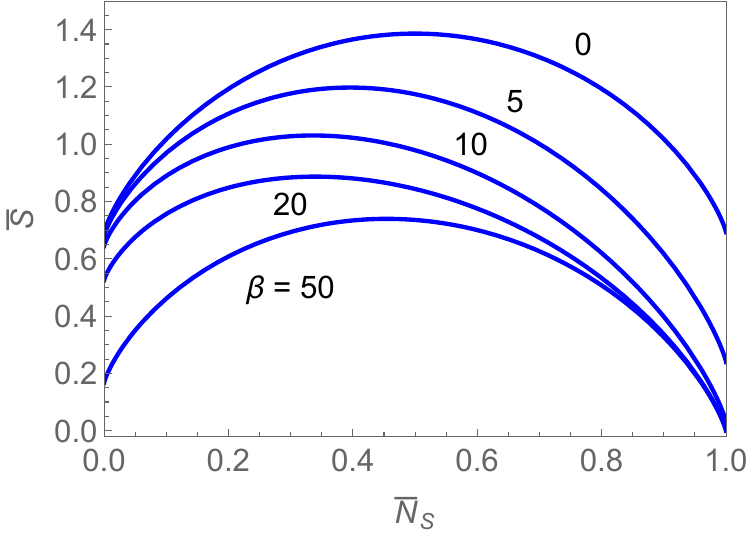}%
\hspace*{5mm}\includegraphics[width=50mm]{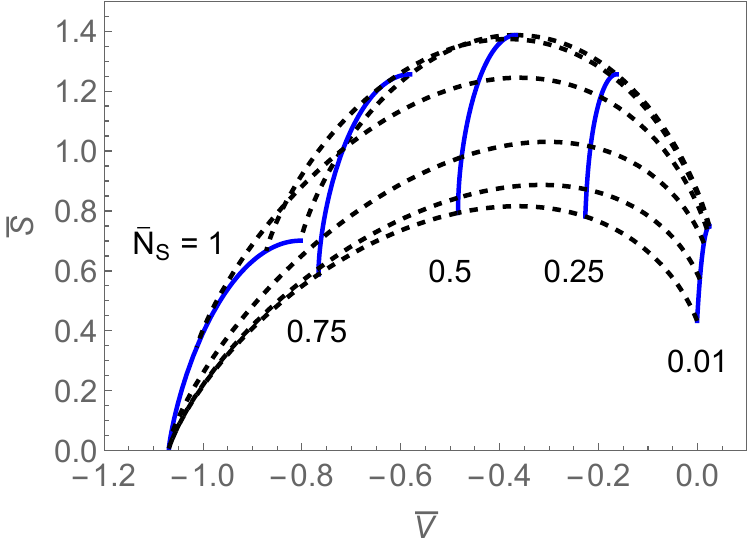}
\end{center}
\caption{Left column: Entropy $\bar{S}$ versus the fraction $\bar{N}_\mathrm{S}$ of small disks for various values of $\beta=1/T_\mathrm{k}$. Right column: The solid curves show $\bar{S}$ versus $\bar{V}$ parametrically for varying $\beta$ at selected values of $\bar{N}_\mathrm{S}$. The dashed curves represent $\bar{S}$ versus $\bar{V}$ for varying $\bar{N}_\mathrm{S}$ at $\beta=0,5,10,20,30$ (top to bottom in each panel on the right). The choice of activation energies favors size segregation with $\Delta\mathcal{V}=0.197\ldots$ from(\ref{eq:31}) (top row), size alternation with $\Delta\mathcal{V}=-0.197\ldots$ (middle row), and size randomness with $\Delta\mathcal{V}=0$ (bottom row).}
  \label{fig:1}
\end{figure}

Jamming protocols are likely to exist which are adequately represented by activation energies (\ref{eq:12}) such that a scenario with $\Delta\mathcal{V}>0$ or a scenario with $\Delta\mathcal{V}<0$ is realized. 
Here we visualize the distinct ordering tendencies in the two scenarios and show that neither ordering tendency prevails in the border scenario, $\Delta\mathcal{V}=0$.
The six panels in Fig.~\ref{fig:1} depict the entropy (\ref{eq:20}) versus the fraction of small disks (\ref{eq:8}) on the left and the same entropy versus the excess volume (\ref{eq:7}) on the right for the three scenarios favoring size segregation (top row), size alternation (middle row), and size randomness (bottom row).

Comparing the panels on the left, we see that the top entropy curve $(\beta=0)$ is the same.
When the disks are jammed from a state of high-intensity agitations, the activation energies have no impact.
There are two contributions to that entropy, $\bar{S}_\mathrm{mix}$ originating in the random mixing of large and small disks, and $\bar{S}_\mathrm{con}$ originating from the mix of configurations accessible to disks of the same size.
For $\bar{N}_\mathrm{S}=0$ or $\bar{N}_\mathrm{S}=1$, only $\bar{S}_\mathrm{con}$ is present, whereas $\bar{S}_\mathrm{mix}$ maximizes at $\bar{N}_\mathrm{S}=\frac{1}{2}$. 
The two parts are not fully separate nor are they additive.

Upon lowering the intensity $T_\mathrm{k}$ of random agitations, the particle activation energies (\ref{eq:12}) as determined by the jamming protocol come into play. 

\begin{itemize}

\item[--] In cases with $\Delta\mathcal{V}>0$ such as the one selected for the top row, the prevailing segregation tendency lowers $\bar{S}_\mathrm{con}$ and $\bar{S}_\mathrm{mix}$ at rates that lower and flatten the curves.
The entropy approaches zero  for all values of $\bar{N}_\mathrm{S}$ in the limit $\beta\to\infty$.

\item[--] In cases with and $\Delta\mathcal{V}<0$ such as the one selected for the middle row, the alternation tendency is prevalent.
Here the lowering of $\bar{S}_\mathrm{mix}$ is, in general, impeded by the surplus of majority disks.
Only for $\bar{N}_\mathrm{S}=\frac{1}{2}$ can this ordering tendency be accommodated into a fully ordered, zero-entropy configuration.

\item[--] In the borderline case $\Delta\mathcal{V}=0$ such as selected for the bottom row, the particle activation energies are balanced such that neither ordering tendency has the edge. 
$\bar{S}_\mathrm{mix}$ stays high. 

\item[--] For $\bar{N}_\mathrm{S}=0$ or $\bar{N}_\mathrm{S}=1$, where only disks of one size are present, the entropy is purely configurational in nature and approaches zero in the limit $T_\mathrm{k}\to0$ as reported earlier \cite{janac1}.

\end{itemize}

The volume of jammed configurations depends on the fraction of small disks and the degree of disorder in the disk configurations. 
There is general tendency of excess volume $\bar{V}$ to diminish as $\bar{N}_\mathrm{S}$ increases and as the available configurations at fixed $\bar{N}_\mathrm{S}$ become more ordered.
The evidence for these observations is best seen in the solid of the panels on the right.
The rate and extent at which compactification establishes itself depends on the the prevailing ordering tendencies.

\section{Conclusion}\label{sec-8}

This study demonstrates a scenario in which size segregation of granular is the consequence of unbiased steric forces alone, with no assistance by external forces or asymmetric shape of confinement. 
In this scenario, the ordering according to size, in the form of segregation or alternation, is a manifestation of symmetry breaking. 
The scenario investigated in this study is admittedly artificial. 
However, this is not unusual for key results based on exact analysis. It suggests that the same effect may be observable in scenarios that are closer to possible experimental realizations.
How exactly our choices of activation energies are related to actual jamming protocols remains an open question. 
It will take a major study of stochastic modeling to work this out.
Nevertheless, we showed that the two opposite ordering tendencies are robust and depend only on the setting of single physical parameter.


\begin{thebibliography}{100}


\bibitem{PHBJ22}
A. V. Patil et al.,
Comp. Part. Mech. \textbf{9}, 693 (2022).

\bibitem{ABS22}
M. I. H. Ansari, A. Bhateja, and I. Sharma,
arXiv: 2210.08040.

\bibitem{BS06}
R. K. Bowles and I. Saika-Voivod,
Phys. Rev. E \textbf{73}, 011503 (2006).

\bibitem{AB09}
S. S. Ashwin and R. K. Bowles,
Phys. Rev. Lett. \textbf{102}, 235701 (2009).

\bibitem{BA11}
R. K. Bowles and S. S. Ashwin,
Phys. Rev. E \textbf{83}, 031302 (2011).

\bibitem{janac1}
N. Gundlach, M. Karbach, D. Liu, and G. M\"uller,
J. Stat. Mech. P04018 (2013).

\bibitem{janac2}	
C. Moore, D. Liu, B. Ballnus, M. Karbach, G. M\"uller,  J. Stat. Mech. P04008 (2014).

\bibitem{janac3}
D. Liu and G. M\"uller,
Phys. Rev. E \textbf{105}, 024904 (2022).

\bibitem{GZB23}
J. M. Gramlich, M, Zarif, an R. K. Bowles,
Soft Matter \textbf{19}, 1373 (2023).

\bibitem{Hald91a}
F. D. M. Haldane, Phys. Rev. Lett. \textbf{67}, 937 (1991).

\bibitem{Wu94}
Y.-S. Wu, Phys. Rev. Lett. \textbf{73}, 922 (1994).

\bibitem{pichs}
D. Liu, J. Vanasse, G. M\"uller, and M. Karbach,
Phys. Rev. E \textbf{85}, 011144 (2012).

\end{thebibliography}
\end{document}